\documentstyle[12pt]{article}
\begin{document}

\overfullrule 0 mm
\language 0

\centerline { \bf{ DELAY EQUATION }}
\centerline { \bf{FOR CHARGED BROWN PARTICLE}}
\vskip 0.5 cm \centerline {\bf{ Alexander A.  Vlasov}}
\vskip 0.3 cm \centerline {{  High Energy and Quantum Theory}}
\centerline {{  Department of Physics}} \centerline {{ Moscow State
University}} \centerline {{  Moscow, 119899}} \centerline {{
Russia}} \vskip 0.3 cm
 {\it 
In previous work ( physics/0004026) was shown,  with the 
help of numerical calculations, that the effective Brown temperature for 
charged particle is lower than that  for particle without charge.   Here 
we derive this result without numerical calculations, integrating the 
delay equation analytically, as for zero, so for nonzero viscosity.  }

03.50.De
\vskip 0.3 cm

{\bf {1.}}

\vskip 0.2 cm
 To describe motion of charged Brown particle in so called 
"extended quasi-stationary approximation"[1] in [2] was used the 
Sommerfeld model [3] of charged rigid sphere. The equation of 
straightline motion of such Brown particle in dimensionless form 
reads [2]:
$$\dot y(x)
=f(x)+\gamma \cdot\left[ y(x-\delta) - y(x)\right] \eqno (1)$$ 
here

$y(x)$ - is dimensionless velocity of the particle;

$x$ - is dimensionless "time";

$f(x)$ - is some external (stochastic) force;

$\delta$ - is  "time" delay;

$\gamma$ - is coefficient: $\gamma \cdot \delta$  is proportional to the 
ratio of particle's electromagnetic mass to the mechanical mass:
$\gamma \cdot \delta = (2/3)(Q^2/a)/(mc^2)$ ( $2a$ - is the size of 
Sommerfeld particle of charge $Q$ and mass $m$);

the viscosity $\Gamma$ of the surrounding medium is zero.
 
In [2] was shown,  with the help of 
numerical calculations, that the effective Brown temperature for charged 
particle is lower than that  for particle without charge. Here we derive 
this result without numerical calculations, integrating the delay equation 
(1) analytically.

With zero initial conditions:

$ y=\dot y=0 $ for $x<0$

dividing the $x$-axis into $\delta$ - intervals $ (i-1)\delta \leq x 
\leq i\delta$, $i=1,\ ...,$

and integrating eq. (1) step by step with boundary conditions 
$y_i(x=i\delta)=y_{i+1}(x=i\delta)$, we finally get the recurrence formula:

for $ (N-1)\delta \leq x \leq N\delta$

$$y(x) =y_N(x)=$$
$$\int_{0}^{x} dz f(z) \exp{\gamma(z-x)} +\gamma 
\int_{(N-2)\delta}^{x-\delta} dz\ y_{N-1}(z) \exp{\gamma(z+\delta-x)}$$
$$+\gamma\sum_{i=1}^{N-2} \int_{(i-1)\delta}^{i\delta} dz\ y_{i}(z) 
\exp{\gamma(z+\delta-x)} \eqno(2)$$
with
$$y_1 (x)= \int_{0}^{x} dz f(z) \exp{\gamma(z-x)},\ \ 0<x\leq\delta$$

 Let's consider one interesting case:

 $f(x)$ for intervals $  (i-1)\delta \leq x \leq i\delta$ is constant and 
 is equal to $f_i$.

 Then the eq.(2) for $x=N\delta\equiv x_N$ yields
                      
 $$y^{*}_N \equiv y_N (x=x_N) ={1 \over \gamma}\sum_{k=1}^{N}\ 
 f_k\left[1-C(N-k;p)\right] \equiv \sum_{k=1}^{N}\ f_k\ D_k \eqno(3)$$
where the function $C(n;p)$ is defined as
$$C(n;p)=\exp{(-p(n+1))} \sum_{m=0}^{n}(p\exp{p})^m\ (n+1-m)^m/(m!);
\eqno(4)$$ here $p\equiv \gamma\delta$.

Function $C(n;p)$ is positive and for sufficiently large $n$ (for ex., 
 $n>20$ for $p=1.0$ ) is equival to ${1 \over 1+p}$:  $$C(n;p)_{n\gg 1}= 
{1 \over 1+p} \to\ \ D_n \approx {p \over (1+p)\gamma} \eqno (5)$$ Thus if 
$f_i= f_0$=const $\forall i$, then from (3,5) we get for $N \gg 1$ $$y^{*}_N 
\approx f_0 N\cdot{p \over (1+p)\gamma} = {f_0 \over 1+p}\ 
x_N$$ in accordance with the exact solution of (1) for $f=f_0$=const:  
$$y(x)={f_0 \over 1+p}\ x$$

Also for $N\gg 1$ one can rewrite (3) in the form
$$y^{*}_N \approx {p\over (1+p) \gamma} \sum_{k=1}^{N} f_k =
{\delta \over (1+p) } \sum_{k=1}^{N} f_k \eqno(6)$$
This result resembles the classical Brown result:
from eq.(4) with $\gamma=0$ one immediately gets
$$y(x)=\int_{0}^{x}f(z) dz, \eqno(7)$$
dividing $x$-interval of integration in (7) into $\delta$- intervals with
$f(x)=f_k$ for $  (k-1)\delta \leq x \leq k\delta$, one can take the 
integral in (7) in the following manner:
$$y(x=x_N)= \delta  \sum_{k=1}^{N} f_k \eqno(8)$$
This result differs from  (6) only in the multiplier ${1 \over (1+p) }$.

Thus one can say that the effect of delay (effect of retardation) for 
eq.(1) reduces to the effect of mass renormalization: $m \to m/(1+p)$,
or consequently to the effect of reduction of the external force:
$$f \to\ f/(1+p) \eqno (9)$$
 This result also says that the reduction of the external force is 
model-independent one, and instead of $\gamma\delta$ one can write the 
classical ratio of self-electromagnetic mass to the mechanical mass $m$ in 
its general form:

$$ \gamma \delta \to {1 \over mc^2} \int d\vec r d\vec {r'}{\rho(\vec r) 
\rho(\vec {r'}) \over |\vec r - \vec {r'}|} \eqno (10)$$
here $\rho$ - is distribution of charge of a particle.

If $f_k, \ \ k=1,...$ - is the range of stochastic numbers with average 
value $f_a$: $<f_k>= f_a$ (here brackets $<>$ denote time average with 
the same definition as in the classical theory of Brownian motion), then 
eq.(3) yields
$$<y^{*}_N>=f_a \sum_{k=1}^{N} D_k\approx f_a x_N/(1+p) \eqno(11)$$
Consequently the dispersion $D$ is
$$D=(y^{*}_N -<y^{*}_N>)^2 = \sum_{k=1}^{N} \sum_{m=1}^{N} D_k D_m <(f_k 
-f_a)(f_m-f_a)>$$
$$ =\sum_{k=1}^{N} \sum_{m=1}^{N} D_k D_m R(k-m) \eqno (12)$$
here $R(k-m)$ - is correlation function of stochastic force $f$.
If $R$ is compact:
$$R(k-m)=R_0 \delta_{mk}/\delta \eqno (13)$$
then the dispersion (12) is 
$$D= R_0/ \delta \sum_{k=1}^{N} (D_k)^2 \approx R_0 x_n/(1+p)^2 \eqno 
(14)$$ This result should be compared with classical one.

The theory of Brownian motion without viscosity tells ( eq. (1) with 
$\gamma=0$ ) that the dispersion $D_B$ is
$$D_B=\int_{0}^{x}dz_1 \int_{0}^{x}dz_2 \cdot R(z_1-z_2) \eqno (15)$$
here $R(z_1-z_2)= <(f(z_1) -f_a)(f(z_2)-f_a)>$ - is the correlation 
function. If
 $$R(z_1-z_2)= R_0 \delta (z_1-z_2) $$
then
$$D_B= R_0 x \eqno (16)$$
Consequently we see that (eqs. (16) and (14) ) the dispersion of the 
Sommerfeld charged particle is lower than that of the classical Brown 
particle without electric charge: $D=D_B(1+p)^{-2}$. Thus one can say that 
the effective temperature of Sommerfeld particle is lower than that of the 
Brown one.  This result is model independent one (see the remark made 
above - eq. (10) ). 

So we confirm the result of the work [2]. 
\eject

\vskip 0.5 cm

{\bf {2.}}

\vskip 0.2 cm

If the viscosity $\Gamma$ is not zero, the main equation reads:
$$\dot y(x) +\Gamma \cdot y(x)
=f(x)+\gamma \cdot\left[ y(x-\delta) - y(x)\right] \eqno (17)$$ 
For $f=f_0= const$ eq.(17) has the exact solution
$$y(x)={f_0 \over \Gamma}(1-\exp{(-ax)}) \eqno(18)$$
and $a$ is determined by the eq.
$$\Gamma+ \gamma -a = \gamma\exp{(a\delta)} \eqno(19)$$
 Iterative 
solution $y(x_N)=y^{*}_N$ of eq.(17), if $ f(x)=f_i=const$ for 
intervals $ (i-1)\delta \leq x \leq i\delta$, can be put in the form:  
 $$y^{*}_N=\sum_{k=1}^{N} f_k D_k \eqno(20)$$ here $D_k$ - some discrete 
 function which can be found from recurrence formula, analogous to (2). 
But it is convenient to find $D_k$ from the following considerations, 
using exact results (18,19).  Solution (20) must tend to the exact 
solution (18) (in the case $f_i=f_0=const \ \forall\  i$) if the $x$-axis 
is divided into infinitesimally small $\delta$-intervals: $\delta \to 0$ 
and $N \to \infty$ in such a way that $x_N=\delta \cdot N =const$.  Thus 
one can rewrite 
$$y^{*}_N=\sum_{k=1}^{N} f_k D_k =f_0 \sum_{k=1}^{N} D_k= 
{f_0 \over \Gamma}(1-\exp{(-ax_N)})$$
 so $$\sum_{k=1}^{N}  D_k=  {1 \over 
\Gamma}(1-\exp{(-a\delta N)}) \eqno (21)$$
 If $\delta \to 0$ we can replace the 
sum in lhs of (21) by the integral: 
 $$\sum_{k=1}^{N} 
D_k \approx \int^{N} D_k dk= {1 \over \Gamma}(1-\exp{(-a\delta N)}) \eqno 
(22)$$ 
Differentiation of (22) with respect to $N$ provides us with this 
expression for $D_N$:

$$D_N \approx {a\delta\over \Gamma}\exp{(-a\delta N)} \eqno (23)$$
Substitution of (23) back into (21) gives

$$\sum_{k=1}^{N}{a\delta\over \Gamma}\exp{(-a\delta k)}= 
{a\delta\over \Gamma}\cdot {1-\exp{(-a\delta N)}\over \exp{(a\delta})-1}\eqno (24)$$
Consequently the required result (rhs of (21) ) is reproduced if we 
expand the denominator in (24) in the following way:
  $$\exp{(a\delta})-1 
\approx a\delta \eqno (24)$$

Using this representation of $D_k$, one can find the dispersion D. For 
correlation function (13) we have
$$D= {R_0 \over  \delta} \sum_{k=1}^{N} (D_k)^2 \approx $$
$${R_0 \over  \delta}\cdot
({a\delta\over \Gamma})^2 \cdot {1-\exp{(-2a\delta N)}\over \exp{(2a\delta})-1} \approx {R_0 a\over 2 (\Gamma)^2}
 (1-\exp{(-2a\delta N)})\eqno (25)$$
here we expanded the expression $\exp{(2a\delta})-1$ in the same manner as 
in (24): $\exp{(2a\delta})-1 \approx 2a\delta$.

Solving the eq.(19) in approximation (24), we find
$$a \approx {\Gamma\over (1+\gamma \delta)} \eqno (26)$$
 So with (26) and (25) the dispersion D takes the form
$$D={R_0 \over 2 (\Gamma)(1+\gamma \delta)} (1-\exp{(-2a\delta N)})\eqno 
(27)$$

Dispersion (27) for $\gamma \equiv 0$ is exactly the same as Brownian 
dispersion $D_B$:
$$D_B={R_0 \over 2\Gamma} (1-\exp{(-2 \Gamma x_N)})$$

If $a x_N \ll 1$, solution (27) yields
$$D \approx {R_0 x_N \over (1+\gamma \delta)^2}$$
i.e. the solution we have got earlier (14).

If $a x_N \gg 1$,  (27) yields
$$D\approx {R_0  \over 2\Gamma (1+\gamma \delta)} = {D_B \over  (1+\gamma 
\delta)}$$  
 Thus dispersion $D$ differs from  the Brownian one. Consequently the 
effective temperature of charged particle, undergoing Brownian motion, is 
lower then that of particle without charge. Now we have proved this result 
in general case of nonzero viscosity. Of course, our general conclusion is 
model-independent one - see the above remark (10).

 \vskip 2 cm \centerline {\bf{REFERENCES}}

  \begin{enumerate}

\item  T.Erber, Fortschr. Phys., 9, 343 (1961).
\item Alexander A.Vlasov, physics/0004026.
\item A.Sommerfeld, Gottingen Nachrichten, 29 (1904), 363 (1904), 201
  (1905).

\end{enumerate}

 \end{document}